\documentclass[runningheads]{svmult}

\usepackage{makeidx}   
\usepackage{graphicx}  
\usepackage{subeqnar}  
\usepackage{multicol}  
\usepackage{physprbb}  
\makeindex             

\newcommand{\eq}[1]{Eq. (\ref{#1})}

\begin{document}
\title*{The $g$ Factor of a Bound Electron\protect\newline in a Hydrogen-Like Atom}
\toctitle{The $g$ Factor of a Bound Electron in a Hydrogen-Like Atom}
\titlerunning{$g$ Factor of Bound Electron}
\author{
Savely G. Karshenboim\inst{1,2}\thanks{E-mail: sek@mpq.mpg.de}
}
\authorrunning{Savely G. Karshenboim}
\institute{
D. I. Mendeleev Institute for Metrology, 
198005 St. Petersburg, Russia
\and Max-Planck-Institut f\"ur Quantenoptik, 
85748 Garching, Germany
}

\maketitle              

\begin{abstract}
Recently, a precise measurement on the bound electron $g$ factor in hydrogen-like carbon was 
performed \protect{\cite{werth}}. 
We consider the present status of the theory of the 
$g$ factor of an electron bound in a hydrogen-like 
atom and discuss new opportunities and possible applications of the recent experiment.
\end{abstract}

\section{Introduction}

Until now any precision test of bound state QED \cite{icap} was always realized in the way that 
only a final theoretical figure could be compared with some experimental result and no 
term of any theoretical expression could be tested separately. Any theoretical expression is a 
function of few parameters and one of them is the nuclear charge $Z$. However, there was no way 
to measure any function of $Z$. Only very few particular values of $Z$ were 
available for precision experiments of the Lamb shift, fine and hyperfine structure. Now this has 
been changed dramatically.

Recently a bound electron $g$ factor in the hydrogen-like carbon ion was measured 
very accurately by the Mainz-GSI collaboration \cite{werth}. We consider here this new 
opportunity to test bound state QED and to determine precisely some fundamental 
constants from the study of the bound electron $g$ factor. 

In contrast to the study of the Lamb shift and hyperfine structure, 
it is possible to perform 
experiments on the $g$ factor of the bound electron in different hydrogen-like ions with about 
the same accuracy. The experiment [1] is now in progress and some other hydrogen-like 
ions can be measured soon. This provides a possibility to learn about 
the bound $g$ factor as a 
function of the nuclear charge $Z$ and the nuclear mass number $A$. The ions under  
study  \cite{werth} must have spinless nuclei and so they have the most simple level 
scheme.

The measured value is equal to $(Z-1)m/gM$, where $m$ is the electron mass, and $M$ is the 
nuclear mass. The latter is usually known accurately in atomic mass units. If we 
determine the bound electron $g$ factor from the theory, then the experiment leads to 
a new precision value of the electron mass in atomic units. The accuracy of the electron 
mass value is compatible with the one from a direct measurement \cite{mass}. 
The uncertainty of the theory of the $g$ factor 
in the carbon ion is on about the same level 
as the experimental one. The theoretical precision can be improved 
if the berillium ion, where the only unknown correction 
$(Z\alpha)^4m/M$ is small and negligible, is studied.

Another way to manage experimentally the problem of the unknown term is to use 
another kind of theory. 
The usual theory in the case of simple atoms (we call it {\em strong theory}) give us some 
values, while in case of the $g$ factor 
it must be useful even to fix a form of the theoretical expression (we call 
this option {\em weak theory}). When one knows part of the theoretical terms and a form 
of other terms, it is possible try to measure the unknown coefficients. The only 
unknown term which is important for carbon and ions with $Z$ around 6 is of order 
$(Z\alpha)^4m/M$. One can take an advantage of the {\em weak theory} approach and measure 
the coefficient of this term by studying with the oxygen ion.

An important feature of the study of 
the $g$ factor of a bound electron at $Z = 20-30$ is also the  
possibility to learn about higher-order two-loop corrections, which are one of the crucial 
problems of bound state QED theory.
Below we discuss in detail the present status of theory 
and experiment. We consider a new opportunity to precisely test bound state QED and to 
accurately determine two fundamental constants: the electron-to-proton mass ratio and the fine 
structure constant.

Below we discuss in detail a number of problems arising due to the 
precision study of the bound electron $g$ factor.

\section{Theory}
\subsection{Present status}
We follow the notation of our paper \cite{PLA} and 
present the $g$ factor of a bound electron in 
a hydrogen-like ion
in the form
\begin{equation}
g_e = 2 \cdot (1 + a + b)
\end{equation}
where the bound electron anomalous magnetic moment at low $Z$ is $b\sim (Z\alpha)^2$, 
while the free part of the anomalous magnetic moment is $a\sim \alpha/2\pi$. 
It is useful to split the bound-state electron anomaly into three terms \cite{PLA}:
\begin{equation}
\label{b1aprime}
b=1\cdot b_1 + a\cdot b_a + b^\prime
\;,
\end{equation}  
where two first terms \cite{kin1,kin2}
\begin{eqnarray}
 b_1 &=& \frac{1+2\sqrt{1-(Z\alpha)^2}}{2}
+(Z\alpha)^2\left[\frac{1}{2}\frac{m}{M}-
\frac{1+Z}{2}\left(\frac{m}{M}\right)^2\right]
\nonumber\\
&\simeq&
(Z\alpha)^2\left[-\frac{1}{3}+\frac{1}{2}\frac{m}{M}-
\frac{1+Z}{2}\left(\frac{m}{M}\right)^2-\frac{1}{12}(Z\alpha)^2\right]
\end{eqnarray}
and \cite{kin1a,kin2}
\begin{equation}
\label{baterm}
b_a=(Z\alpha)^2\left[\frac{1}{6}-\frac{1}{3}\frac{m}{M}\right]
\end{equation}
are due to kinematic effects, while the last term is due to some 
non-trivial bound-states effects. The $b^\prime$ term starts from the order of 
$(Z\alpha)^4$ and contains QED ($\alpha(Z\alpha)^4m$ and higher) \cite{BCS,PSSL,beier},  
nuclear size ($(Z\alpha)^4(mR_N)^2$ and higher) \cite{PSSL} and recoil 
($(Z\alpha)^4m/M$ and higher) corrections.  
In Table 1 we present the most accurate published values \cite{PSSL} 
of different contributions to $b^\prime$. The uncertainty 
was not specified there and we estimate it as half the value of the last digit. More precise 
calculations are presented in Ref. \cite{beier}.
\begin{table}
\caption{Some contributions to $b^\prime$ in units of 
$10^{-9}$ {\rm \protect{\cite{PSSL}}} due to  
vacuum polarization ($VP$), self-energy ($SE$) and finite nuclear 
size ($NS$). The nuclear size effects were studied there 
for the main isotope of each element}
\label{Tab1}
\begin{center}
\def\arraystretch{1.4}
\setlength\tabcolsep{5pt}
\begin{tabular}{r@{\hskip 3em}r@{\hskip 3em}r@{\hskip 3em}r}
\hline
$Z$ & $b^\prime_{\rm VP}~~~~$ & $b^\prime_{\rm SE}~~~~~$ & $b^\prime_{\rm NS}~~~$ \\
\hline
~1  & -0.00(3)       & 0.08(3)        &            \\
~2  & -0.05(3)       & 0.93(3)        &            \\
~4  & -0.86(3)       & 12.77(3)       &            \\
~6  & -4.2(3)~       & 55.2(3)~       &            \\
~8  & -13.2(3)~      & 151.4(3)~      &  0(7)~        \\       
10  & -31.8(3)~      & 327.1(3)~      &  0(7)~        \\
12  & -64.6(3)~      & 610.9(3)~      &  5(7)~        \\
14  & -118(3)~~~     & 1\,030(3)~~~     &  10(7)~       \\
16  & -200(3)~~~     & 1\,610(3)~~~     &  20(7)~       \\
18  & -309(3)~~~     & 2\,381(3)~~~     &  35(7)~       \\
24  & -948(3)~~~     & 6\,127(3)~~~     &  135(7)~      \\
32  & -2\,890(3)~~~    & 16\,808(30)~~    &  620(70)     \\
\hline
\end{tabular}
\end{center}
\end{table}

\subsection{Comparison to the experiment}

Precision studies of the $g$ factor of a bound electron in simple atoms 
 were started with experiments on
hydrogen \cite{h} and its comparison with deuterium 
\cite{d1,d2,d3} and tritium \cite{t}, as well as on the helium ion \cite{he}.
In particular in case of low $Z$ the result for the non-trivial bound-state 
term 
\begin{equation}
b^\prime(Z=1) = 0.1\cdot10^{-9}
\end{equation}
and
\begin{equation}
b^\prime(Z=2) = 0.9\cdot10^{-9}
\end{equation}
is small enough and that explains why experimental results from the precision measurements 
done decadse ago have been out of the theoretical interest for two last decades. 

We summarize experimental and theoretical data in Table 2. Some of them were obtained indirectly 
and we present in Appendix Table \ref{TabA} all important auxiliary measurements. 
\begin{table}
\caption{Comparison of theory and experiment. Here, $g_0(e)$ stands for the magnetic 
moment of a free electron and it contains the anomalous magnetic moment $a\sim \alpha/2\pi$}
\label{Tab2}
\begin{center}
\def\arraystretch{1.4}
\setlength\tabcolsep{5pt}
\begin{tabular}{c@{\hskip 3em}lc@{\hskip 3em}l}
\hline
Value & Experiment & Reference & Theory\\
\hline
$ \frac{g({\rm H})}{g_0(e)} $ 
     & $1- 17.709(13)\cdot10^{-6} $ & \protect{\cite{h,d1}} 
         & $1- 17.694\cdot10^{-6} $ \\
$ \frac{g({\rm D})}{g({\rm H})} $ 
     & $ 1-7.22(3)\cdot10^{-6} $ & \protect{\cite{d3}}
         & $ 1-7.24\cdot10^{-6} $\\
$ \frac{g({\rm T})}{g({\rm H})} $ 
     & $ 1-10.7(15)\cdot10^{-6} $ & \protect{\cite{t}} 
         &$ 1-9.7\cdot10^{-6} $ \\
$ \frac{g({}^4{\rm He}^+)}{g_0(e)} $ 
     & $1- 70.88(30)\cdot10^{-6} $ & \protect{\cite{d1,h,he,rb87,rb85,8785}}
         & $1- 70.91\cdot10^{-6} $ \\
$ \frac{g({}^{12}_{~6}{\rm C}^{5+})}{2} $ 
     & $1+ 520.798(2)\cdot10^{-6} $ & \protect{\cite{werth}}
         & $1+ 520.795(1)\cdot10^{-6} $ \\
\hline
\end{tabular}
\end{center}
\end{table}

In the case of the recent experiment with hydrogen-like carbon the non-trivial QED effects 
contribute an observable amount (see Table 1). We need to mention that, due to some 
delay of 
the final publications of the experimental result \cite{werth} and theoretical calculations 
\cite{beier}, no actual theoretical predictions have been published. Most of the presentations 
(conference and seminar talks and posters) dealt with unaccurate theoretical predictions  
because it was believed that nothing had been known on the two-loop corrections. However, 
that was not the case, because from the beginning of the theoretical calculations  
up to recent re-calculations it was clearly stated ed \cite{kin1a} that the $(Z\alpha)^2$ term  
in \eq{baterm} is of pure kinematic origin and so the result
is valid in any order of the expansion in $\alpha$ 
for the anomalous magnetic moment of a free electron, and in particular
\begin{eqnarray}
\Delta b({\rm two-loop})&\simeq&\left[-0.328...\left(\frac{\alpha}{\pi}\right)^2\right]
\left[\frac{1}{6}(Z\alpha)^2\right]\nonumber\\
& =& - 5.7\cdot10^{-10}\;.
\end{eqnarray}
 We also confirmed this statement in our 
paper \cite{PLA}. Our use of the known result on the $\alpha^2(Z\alpha)^2$ term leads to 
\begin{equation}
b({}^{12}_{~6}{\rm C}^{5+}) = 1+ 520\,795(1)\cdot10^{-9}
\end{equation}
and offers the accuracy presented in Table 2.

\subsection{Potential model}

The bound $g$ factor is now a new value that must to be precisely calculated and 
it may be useful to compare the calculation of the $g$ factor with 
evalutions for the energy levels. 
A number of corrections can be described with the help of an effective potential, 
and in particular effects due to vacuum polarization of free electrons, 
the Wichmann-Kroll potential and nuclear size can be studied in the leading 
non-relativistic approximation with the use of a delta-like potential
\begin{equation}
\label{vda}
V({\bf r}) = A \delta({\bf r})\;,
\end{equation}
where the energy shift of the ground state due to this potential is of the 
form\footnote{We use through the paper the relativistic units in which $\hbar=c=1$.} 
\begin{equation}
\label{eda}
\Delta E_A = A \, \frac{(Z \alpha m)^3}{\pi}
\end{equation}
and for different corrections the shift is 
\begin{eqnarray}
\label{aaa}
\Delta E_{VP} &=&-\frac{4}{15}\frac{\alpha(Z \alpha )^4m}{\pi}\;, 
\nonumber\\
\Delta E_{WK} &=& \left(\frac{19}{45}-\frac{\pi^2}{27}\right)
\frac{\alpha(Z \alpha )^6m}{\pi}\;,
\nonumber\\
\Delta E_{NS} &=&\frac{2}{3}\,\frac{(Z \alpha )^4m }{\pi}\,(mR_N)^2\;.
\end{eqnarray}

We have calculated a correction to the bound electron anomaly and 
the result is \cite{PLA} 
\begin{equation}
\label{baea}
\Delta b^\prime_{A}=  2\,\frac{\Delta E_{A}}{m}\;.
\end{equation}
In particular, for the finite-nuclear-size correction one obtains
\begin{eqnarray}
\label{fns}
\Delta b^\prime_{NS} &=& \frac{4}{3}\, (Z \alpha )^4 \,(mR_N)^2
\nonumber\\
&\simeq& 2.5355\cdot 10^{-14}\,Z^4\,{\cal R}^2\;,
\end{eqnarray}
where the value root-mean-square charge radius ${\cal R}$ in fermi for carbon $^{12}$C is  
equal to 2.478(6) \cite{RC12} and 
\begin{equation}
\Delta b^\prime_{NS} (^{12}{\rm C}^{5+}) =  0.2\cdot 10^{-9}\;.
\end{equation}

Our results are in agreement with direct numerical calculations for several corrections 
\cite{PSSL} (see Tables 3 and 4). 
\eq{baea} can be used to estimate unknown corrections if we know the energy 
shift. Actually we put in Tables\footnote{We corrected some misprints in Table 4 (cf. \cite{PLA}) and 
re-estimated the uncertainty for nuclear-size correction with the result 
$\delta b^\prime_{NS}=\Delta b^\prime_{NS}\cdot (Z\alpha)^2\ln(Z\alpha m R_N)$.} 3 and 4
the uncertainty of the analytic results found in this way \cite{PLA}.
However, one has to remember that the 
theory of the bound $g$ factor is more complicated than the one for the Lamb 
shift and sometimes the bound anomaly can be more logarithmic. An example is the order 
$\alpha^2(Z\alpha)^4$: the $b^\prime$ term contains a logarithm of $(Z\alpha)$ and a non-analytic 
contribution due to low energy effects, 
while the contribution to the Lamb shift in this order has no logarithm and 
completely originates from the high energy contribution. Higher-order two-loop terms have not 
been evaluated yet even in the case of the energy levels of low-$Z$ atoms, and we do not anticipate 
the results in near future. As we mentioned, the bound $g$ factor involve more complicated calculations 
than the theory of energy levels, and a reason of our study using effective non-relativistic technics 
is to find an appropriate approach for effective $(Z\alpha)$ expansions which can be used to study 
two-loop corrections.
\begin{table}
\caption{Comparison of the vacuum polarization contribution:  
\protect{\eq{baea}} against \protect{\cite{PSSL}}. The uncertainty of the analytic result 
was discussed in Ref. \protect{\cite{PLA}} }
\label{Tab3}
\begin{center}
\def\arraystretch{1.4}
\setlength\tabcolsep{5pt}
\begin{tabular}{r@{\hskip 3em}r@{\hskip 3em}r}
\hline
$Z$ & Numerical result &   Analytic result\\
     & Ref. \protect{\cite{PSSL}}~~~~ & \protect{\eq{baea}}~~~~\\
& $[10^{-9}]$~~~~ & $[10^{-9}]$~~~~\, \\[1ex]
\hline
~1  & -0.00(1)~   & -0.003\,5\,\,     \\
~2  & -0.05(1)~   & ~~~-0.056(1)       \\
~4  & -0.86(1)~   & -0.90(3)~       \\
~6  & -4.2(1)~~   & -4.6(2)~~~      \\
~8  & -13.2(1)~~  & -14(1)~~~~~    \\         
10  & -31.8(1)~~  & -35(3)~~~~~   \\
12  & -64.6(1)~~  & -73(8)~~~~~   \\
14  & -118(1)~~~~ & -135(17)~~~~ \\
16  & -200(1)~~~~ & -230(33)~~~~ \\
18  & -309(1)~~~~ & -369(59)~~~~  \\
20  &            & -560(100)~~~ \\
22  &            & -830(160)~~~ \\
24  & -948(1)~~~~ & -1\,170(250)~~~  \\
26  &            & -1\,600(370)~~~  \\
28  &            & -2\,160(540)~~~ \\
30  &            & -2\,850(760)~~~ \\
32  & -2\,890(1)~~~~ &  -3\,680(1\,060)\,\,~  \\
\hline
\end{tabular}
\end{center}
\end{table}
\begin{table}
\caption{Finite-nuclear-size contribution: 
\protect{\eq{baea}} against \protect{\cite{PSSL}}. The uncertainty of the analytic result 
was discussed in Ref. \protect{\cite{PLA}} }
\label{Tab4}
\begin{center}
\def\arraystretch{1.4}
\setlength\tabcolsep{5pt}
\begin{tabular}{rl@{\hskip 3em}r@{\hskip 3em}r}
\hline
$Z$ & ~~$R_N$  &  Numerical result & Analytic result\\
&    & Ref. \protect{\cite{PSSL}}~~~~~~ & \protect{\eq{baea}}~~~~~ \\
&  ~~[fm] & $[10^{-9}]$~~~~~~ & $[10^{-9}]$~~~~~~ \\[1ex]
\hline
~8  & 2.737(8)     &  0(3)~~~~~    &  0.78(6)~~~  \\         
10  & 2.992(8)     &  0(3)~~~~~    &  2.27(9)~~~  \\
12  & 3.08(5)      &  5(3)~~~~~    &  5.0(3)~~~~  \\
14  & 3.086(18)    &  10(3)~~~~~   &  9.2(7)~~~~  \\
16  & 3.230(5)     &  20(3)~~~~~   & 17(2)~~~~~~  \\
18  & 3.423(14)    &  35(3)~~~~~   & 31(4)~~~~~~ \\
24  & 3.643(3)     &  135(3)~~~~~  & 112(22)~~~~~ \\
32  & 4.088(8)     &  620(3)~~~~~  & 444(145)~~~~ \\
\hline
\end{tabular}
\end{center}
\end{table}

\section{Precise tests of the bound state QED}

Now let us discuss possible precision tests of bound state QED. 
First we need to discuss what experimental results have been available up-to-now \cite{icap}:
\begin{itemize}
\item Precision measurements of the Lamb shift are possible for $Z=1,2$. 
\item The fine structure can be precisely studied for $Z=1$ and $Z=6$. 
\item Transitions in the gross structure can be accurately measured at $Z=1,2$.
\item The hyperfine structure of the 1s and 2s levels can be obtained precisely for the 
hydrogen atom and its isotops\footnote{The tritium 2s {\em hfs} interval 
has not been measured.} and for the helium-3 ion.
\item Precise measurements are also possible for a few transitions in helium.
\item In the case of some high-$Z$ ions, precision experiments are also possible.
\end{itemize}  
Now we list the accurate theoretical predictions that have been available \cite{icap} up-to-date:
\begin{itemize}
\item Accurate calculations for the Lamb shift and {\em hfs} of hydrogen-like atoms 
are limited by their nuclear structure and higher-order QED corrections. In the case 
of low-$Z$ Lamb shift, the finite-nuclear-size effects can be taken into account easily 
if we know the nuclear charge radius.
\item In the case of some high-$Z$ experiments the uncertainty due to the nuclear 
structure disturbs any interpretations in terms of a test of quantum electrodynamics. 
However, the QED calculations are still important for the study of  
nuclear structure. 
\item Few-electron atomic systems cannot be precisely calculated at low $Z$, except some 
special cases (like e. g. isotopic shift). 
\item In the case of high-$Z$ few-electron ions, one of theoretical problems is the higher-order 
electron-electron interaction.
\end{itemize} 

Summarizing this section we need to underline 
that only a few of values of $Z$ are available for the precision 
experiment, a comparison of experiment and theory can be done only for the final values and 
there is a problem of theoretical uncertainty due to unknown higher-order corrections, 
which cannot  be calculated at the moment. We call a theory, which predicts only an final 
value to be compared with an experiment, a {\em strong theory}.

\subsection{Weak and strong theory}

The study of the bound electron $g$ factor provides another option. It is possible now 
to check a {\em part} of the theoretical expression. The expression is a function of the 
nuclear charge ($Z$), the atomic mass ($A$) and the nuclear charge radius ($R_N$). The nuclear 
effects can be easily calculated in the leading order if the nuclear radius is known. 
Since it is possible to measure the $g$ factor for different values of $Z$ and $A$, 
the complete functions can be now available. We put in Fig. 1 all spinless 
nuclei with a not too high value of $Z$ for hydrogen-like ($Z^*=Z$) and 
lithium-like ions ($Z^*=Z-2$). 
The use of ions with different values of the charge-to-mass ratio can be also helpful 
to learn about systematic errors. Due to that it seems to be important to understand which 
progress in calculations for light and medium lithium-like ions is possible in near future.
\begin{figure}[b]
\begin{center}
\includegraphics[width=.8\textwidth]{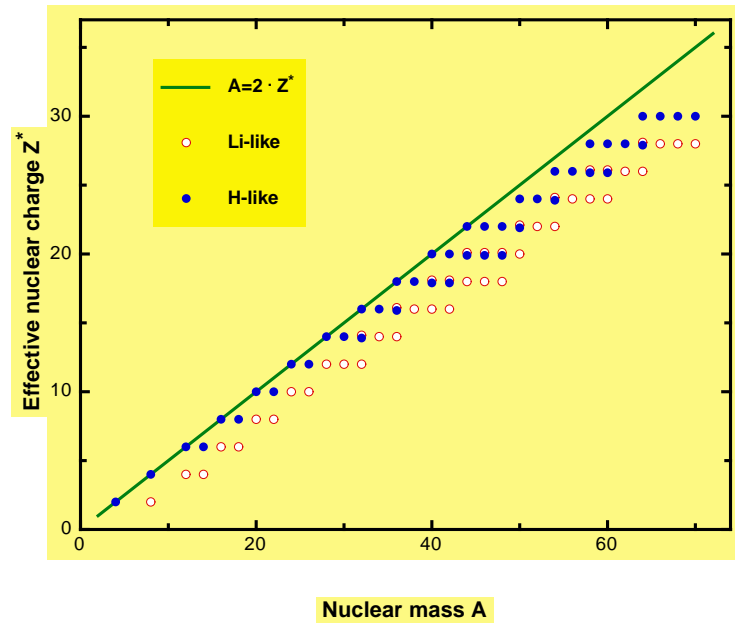}
\end{center}
\caption[]{Hydrogen-like and lithium-like ions with spinless nuclei}
\label{eps1}
\end{figure}

Such a situation with a broad range of available values of 
$Z$ and $A$ offers a different way to compare theory and experiment. It is possible  
now to compare a theoretical expression, which is known only {\em partially}, but with a 
{\em form} of unknown coefficients which is completely understood. 
For example one can write for $Z\sim 6$
\begin{equation}
\label{weakC}
g_e(Z)=g_{\rm known}(Z)+C\,(Z\alpha)^4\frac{m}{M}\;,
\end{equation}
where the first term is known.
Studying experimentally carbon and oxygen the unknown  coefficient $C$ can be determined. 
We call this way to use theory a {\em weak theory}.

\subsection{Higher-order two-loop contributions}

When we go to a higher value of $Z$ ($Z\sim 20$) a wide set of unknown terms have to be taken 
into account. First we need to study all coefficients in equation
\begin{eqnarray}
\label{weakCa}
C_0\,(Z\alpha)^4\frac{m}{M}
&+&C_1\,\alpha(Z\alpha)^4\ln(Z\alpha)+C_2\,\alpha(Z\alpha)^4\nonumber\\
&+&C_3\,(Z\alpha)^5\frac{m}{M}\ln(Z\alpha)+C_4\,(Z\alpha)^5\frac{m}{M}
+C_5\,\alpha^2(Z\alpha)^5\;.
\end{eqnarray}
These coefficients can be studied at $Z=10-20$ or calculated. With $Z$ higher than 20 it is 
necessary to take into consideration $\alpha^2(Z\alpha)^6$ terms which can contain temrs 
up to the cube 
of the low-energy 
logarithm ($\ln(Z\alpha)$) and we have a problem of {\em higher-order two-loop} corrections. 
That is now one of the most important theoretical problems in bound-state QED. In particular, 
it essentially limits computational accuracy for
\begin{itemize}
\item the hydrogen Lamb shift, fine and gross structure;
\item the hyperfine structure intervals in hydrogen, deuterium and especially in the $^3$He$^+$ ion 
(namely, some special combination of that for 1s and 2s levels \cite{ks});
\item the Lamb shift and gross structure in the helium-4 ion;
\item the fine structure in hydrogen-like nitrogen;
\item high-$Z$ few electron atoms (rather ``exact'' calculation in $(Z\alpha)$ is needed).
\end{itemize} 
In the case of the bound $g$ factor for the first time there is an opportunity to study 
the {\em higher-order two-loop} corrections experimentally in detail with varying value of $Z$.

\subsection{Physics of medium $Z$}

The possibility of the direct experimental study of 
these higher-order two-loop terms shows  
the importance of physics of moderate $Z$. Medium-$Z$ theory provides a possibility to 
develop both high-$Z$ and low-$Z$ approaches. In case of low-$Z$ technics, one can 
expand over $(Z\alpha)$ and see if our assumptions on higher-order terms is appropriate or not. 
Using high-$Z$ methods one can perform a $1/Z$ expansion and treat the {\em electron-electron}
interaction in few-electron ions as a perturbation. The study of the $g$ factor of 
lithium-like ions offers an experimental test of present ideas on 
how large higher-order  corrections
electron-electron interaction can be.

\section{Carbon and calcium experiments}

\subsection{Carbon experiment and electron-to-proton mass ratio}

Comparison of theory to precision experiment often involves some other experimental data from 
different fields. In particular, the 
$g$ factor experiment \cite{werth} deals with a comparison of two frequencies: the Larmor spin 
precession frequency 
\begin{equation}
\omega_{\rm L} = g_e \,\frac{e}{2\,m\,c}\, B
\end{equation}
 and the ion cyclotron one 
 \begin{equation}
\omega_{\rm c} = \frac{(Z-1)\,e}{2\,M_I\,c}\, B\;.
\end{equation}
 They are proportional 
to the magnetic field $B$, 
but their ratio is field-independent. Eventually, the experiment leads to a value of $g_e M_I/m$, 
where $M_I$ is the ion mass. There are three sources of 
uncertainty involved: 
\begin{itemize}
\item an experemental for the frequencies $\omega_{\rm L}$ and $\omega_{\rm c}$;
\item a computational one for the $g$ factor; 
\item the one due to the determination of the electron 
mass $m$ in proper units (either in atomic mass units or in terms of the proton mass).
\end{itemize}
 All of them 
are comparable \cite{werth}. The use of theoretical values  
 and the experimental 
data \cite{werth} for the frequencies leads to a value of the electron mass. 

One source of theoretical uncertainty is the unknown term in order $(Z\alpha)^4m/M$ 
(see Eq. (\ref{weakC})) and it might 
be helpful to study $^4$He$^+$ and $^{10}$Be (the correction is negligible) or $^{16}$O and $^{18}$O 
(the correction is 2.4 and 2.1 times larger than that for $^{12}$C). 
It is important to mention that comparing the results on the 
$g$ factor at different $Z$ values, one can check the consistency of the theory without using 
any data of the electron mass. 
Varying $Z$ ($Z=2,4,6,,8$) one can first fix the only unknown coefficient 
and then determine the electron mass.

A study of the electron mass is now of interest also because of determination of the fine structure 
constant $\alpha$ from the photon-recoil-spectroscopy \cite{chu,pritchard}. A measurement of the 
recoil frequency shift
\begin{equation}
{\delta \nu\over \nu} = \frac{h\nu}{2M_Ac^2}
\end{equation}
where $\nu$ is a transition frequency and $M_A$ is an atomic mass, together with a known 
value of frequency $\nu$ gives a value of $h/M_A$. The latter can be compared with the 
Rydberg constant
\begin{equation}
Ry = \alpha^2\,\frac{mc }{h}
\end{equation}
and that allows to determine the fine structure constant $\alpha$. 
The Rydberg constant is known very accurately 
\cite{Ry}, the recoil shift ${\Delta\nu/\nu}$ was measured for $D_1$ cesium line 
in Ref. \cite{chu}, while the frequency of the $D_1$ line  
was determined in Ref. \cite{udem}. To extract $\alpha$ we also need to know $m/M_{Cs}$. The 
cesium mass in atomic mass units was obtained in Ref. \cite{pritchard}. The accuracy of the 
frequency recoil shift \cite{chu} is larger, but compatible with the one for the electron mass 
in atomic units \cite{mass}. It seems \cite{chu} that in near future a determination of the fine 
struture constant from the recoil spectroscopy will be limited by the knowledge of 
the electron mass.

\subsection{Calcium measurement: why it is getting interesting}

Another way to obtain 
the fine structure constant $\alpha$ via the bound electron $g$ factor is to work at higher $Z$.
As we mention, the investigation at $Z\sim 20$ and higher can be useful 
to learn about higher-order two-loop 
corrections. However, the study offers also a new way to determine the fine structure constant. 
For the calcium ion 
($Z = 20$) the bound anomaly $b$ is larger than the free term $a$ and hopefully it is still 
possible to use an expansion in $(Z\alpha)$. A value of $\alpha$ is to be extracted from $(Z\alpha)^2$ 
term.
If a routine work from experimental and theoretical sides (i. e. a measurement in the hydrogen-like 
calcium ion with an uncertainty on a level of $10^{-9}$ and a computation of QED corrections to the 
$g$ factor, which are similar to the ones evaluated some time ago for the energy levels) a value of $\alpha$ 
with uncertainty about $10^{-7}$ can be reached. We expect that an experimental progress is possible 
and further improvement can be expected.

\section{Summary}

Concluding our consideration we would like to underline, that 
the study of the $g$ factor of a bound electron \cite{werth} offers a new opportunity for us to 
precisely test bound state QED theory and to determine two important fundamental constants: 
the fine structure constant $\alpha$ and the electron-to-proton mass ratio $m/m_p$. 
The experiment can be performed 
at any $Z$ with about the same accuracy \cite{werth} and one can expect new data at medium 
$Z$ which will allow to verify 
the present ability to estimate unknown  higher-order corrections (i. e. theoretical uncertainty) 
in both: low-$Z$ and high-$Z$ calculations.

\section*{Aknowledgements}

I am grateful to T. Beier, H. H\"affner, N. Hermanspahn, V. G. Ivanov,
W. Quint, H.-J. Kluge, V. M. Shabaev, J. Reichert, and 
especially to G\"unter Werth for stimulating discussions. 
The work was supported in part by RFBR grant 00-02-16718, NATO grant CRG 960003 and the
Russian State Program ``Fundamental Metrology''. 

\clearpage

\appendix

\section*{Appendix: Auxiliary data on the bound electron $g$ factor measurements}

\begin{table}
\caption{Experimental data for the $g$ factor 
of the hydrogen isotopes and the helium ion}
\label{TabA}
\begin{center}
\def\arraystretch{1.4}
\setlength\tabcolsep{5pt}
\begin{tabular}{c@{\hskip 3em}l@{\hskip 3em}c}
\hline
Value & Result & Reference\\
\hline
$ \frac{g({}^{87}{\rm Rb},\,5S_{1/2})}{g({\rm H,\,1s})} $ 
     & $1+23.585\,5(6)\cdot10^{-6}  $  
         &  \protect{\cite{d1}}\\
$ \frac{g({}^{87}{\rm Rb},\,5S_{1/2})}{g({\rm D,\,1s})} $ 
     & $ 1+23.592\,7(10)\cdot10^{-6} $ 
         &  \protect{\cite{d1}}\\
$\frac{g({}^{87}{\rm Rb},\,5S_{1/2})}{g_0}$ 
     & $1+5.876(13)\cdot 10^{-6}$ 
         & \protect{\cite{h}}\\
$\frac{g({}^4{\rm He}^+,\,1s)}{g({}^4{\rm He},\,2^3S_1)}  $ 
     & $1-29.95(30)\cdot10^{-6}  $ 
         &  \protect{\cite{he}}\\
$ \frac{g({}^4{\rm He},\,2^3S_1)}{g({}^{87}{\rm Rb},\,5S_{1/2})} $ 
     & $1-46.798(50)\cdot10^{-6}  $ 
         &  \protect{\cite{rb87}}\\
$ \frac{g({}^4{\rm He},\,2^3S_1)}{g({}^{85}{\rm Rb},\,5S_{1/2})} $ 
     & $ 1-46.77(7)\cdot10^{-6}  $ 
         &  \protect{\cite{rb85}}\\
$ \frac{g({}^{85}{\rm Rb},\,5S_{1/2})}{g({}^{87}{\rm Rb},\,5S_{1/2})} $ 
     & $ 1-0.000\,41(60)\cdot10^{-6} $ 
         &  \protect{\cite{8785}}\\
\hline
\end{tabular}
\end{center}
\end{table}

\end{document}